\documentclass[preprint]{aastex63}
\usepackage{lineno}

\newcommand{\sm}{$M_\odot$}
\newcommand{\sL}{$L_\odot$}
\newcommand{\tbol}{$T_{\rm bol}$}

\newcommand{\iras}{IRAS 15398$-$3359}

\newcommand{\hhco}{H$_{2}$CO}

\newcommand{\ccchh}{c-C$_3$H$_2$}

\newcommand{\meta}{CH$_3$OH}

\newcommand{\mf}{CH$_3$OCHO}

\newcommand{\nhhcho}{NH$_2$CHO}

\newcommand{\kms}{km s$^{-1}$}
\newcommand{\mjybeam}{mJy beam$^{-1}$}
\newcommand{\metd}{CH$_2$DOH}

\received{}
\revised{\today}
\accepted{}
\submitjournal{ApJ}

\shorttitle{}
\shortauthors{Okoda et al.}

\begin{document}

\title{FAUST VII. Detection of A Hot Corino in the Prototypical Warm Carbon-Chain Chemistry Source IRAS 15398$-$3359}

\correspondingauthor{Yuki Okoda}
\email{yuki.okoda@riken.jp}
\author{Yuki Okoda}
\affiliation{RIKEN Cluster for Pioneering Research, 2-1, Hirosawa, Wako-shi, Saitama 351-0198, Japan}

\author{Yoko Oya}
\affiliation{Center for Gravitational Physics and Quantum Information, Yukawa Institute for Theoretical Physics, Kyoto University, Kyoto, 606-8502, Japan}
\affiliation{Department of Physics, The University of Tokyo, 7-3-1, Hongo, Bunkyo-ku, Tokyo 113-0033, Japan}

\author{Logan Francis}
\affiliation{NRC Herzberg Astronomy and Astrophysics, 5071 West Saanich Road, Victoria, BC, V9E 2E7, Canada}
 \affiliation{Department of Physics and Astronomy, University of Victoria, Victoria, BC, V8P 5C2, Canada}

\author{Doug Johnstone}
\affiliation{NRC Herzberg Astronomy and Astrophysics, 5071 West Saanich Road, Victoria, BC, V9E 2E7, Canada}
\affiliation{Department of Physics and Astronomy, University of Victoria, Victoria, BC, V8P 5C2, Canada}

\author{Cecilia Ceccarelli}
\affiliation{Univ. Grenoble Alpes, CNRS, IPAG, 38000 Grenoble, France}

\author{Claudio Codella}
\affiliation{INAF, Osservatorio Astrofisico di Arcetri, Largo E. Fermi 5, I-50125, Firenze, Italy}
\affiliation{Univ. Grenoble Alpes, CNRS, IPAG, 38000 Grenoble, France}

\author{Claire J. Chandler}
\affiliation{National Radio Astronomy Observatory, PO Box O, Socorro, NM 87801, USA}

\author{Nami Sakai}
\affiliation{RIKEN Cluster for Pioneering Research, 2-1, Hirosawa, Wako-shi, Saitama 351-0198, Japan}

\author{Yuri Aikawa}
\affiliation{Department of Astronomy, The University of Tokyo, 7-3-1 Hongo, Bunkyo-ku, Tokyo 113-0033, Japan}

\author{Felipe O. Alves}
\affiliation{Center for Astrochemical Studies, Max-Planck-Institut f\"{u}r extraterrestrische Physik (MPE), Gie$\beta$enbachstr. 1, D-85741 Garching, Germany}

\author{Eric Herbst}
\affiliation{Department of Chemistry, University of Virginia, McCormick Road, PO Box 400319, Charlottesville, VA 22904, USA}

\author{Mar\'{i}a Jos\'{e} Maureira}
\affiliation{Center for Astrochemical Studies, Max-Planck-Institut f\"{u}r extraterrestrische Physik (MPE), Gie$\beta$enbachstr. 1, D-85741 Garching, Germany}

\author{Mathilde Bouvier}
\affiliation{Leiden University, PO Box 9513, 23000 RA Leiden, The Netherlands}

\author{Paola Caselli}
\affiliation{Center for Astrochemical Studies, Max-Planck-Institut f\"{u}r extraterrestrische Physik (MPE), Gie$\beta$enbachstr. 1, D-85741 Garching, Germany}

\author{Spandan Choudhury}
\affiliation{Center for Astrochemical Studies, Max-Planck-Institut f\"{u}r extraterrestrische Physik (MPE), Gie$\beta$enbachstr. 1, D-85741 Garching, Germany}

\author{Marta De Simone}
\affiliation{European Southern Observatory, Karl-Schwarzschild-Strasse 2, 85748, Garching bei M\"{u}nchen, Germany}
\affiliation{INAF, Osservatorio Astrofisico di Arcetri, Largo E. Fermi 5, I-50125 Firenze, Italy}

\author{Izaskun J\'{i}menez-Serra}
\affiliation{Centro de Astrobiolog\'{\i}a (CSIC-INTA), Ctra. de Torrej\'on a Ajalvir, km 4, 28850, Torrej\'on de Ardoz, Spain}

\author{Jaime Pineda}
\affiliation{Center for Astrochemical Studies, Max-Planck-Institut f\"{u}r extraterrestrische Physik (MPE), Gie$\beta$enbachstr. 1, D-85741 Garching, Germany}

\author{Satoshi Yamamoto}
\affiliation{Department of Physics, The University of Tokyo, 7-3-1, Hongo, Bunkyo-ku, Tokyo 113-0033, Japan}
\affiliation{Research Center for the Early Universe, The University of Tokyo, 7-3-1, Hongo, Bunkyo-ku, Tokyo 113-0033, Japan}




\begin{abstract}
\par We have observed the low-mass protostellar source, IRAS 15398$-$3359, at a resolution of 0\farcs2-0\farcs3, as part of the Atacama Large Millimeter/Submillimeter Array Large Program FAUST, to examine the presence of a hot corino in the vicinity of the protostar. We detect nine CH$_3$OH lines including the high excitation lines with upper state energies up to 500 K. The CH$_3$OH rotational temperature and the column density are derived to be 119$^{+20}_{-26}$ K and 3.2$^{+2.5}_{-1.0}\times$10$^{18}$ cm$^{-2}$, respectively. The beam filling factor is derived to be 0.018$^{+0.005}_{-0.003}$, indicating that the emitting region of \meta\ is much smaller than the synthesized beam size and is not resolved. The emitting region of three high excitation lines,  18$_{3,15}-18_{2,16}$, A ($E_u=$447 K), 19$_{3,16}-19_{2,17}$, A ($E_u=$491 K), and 20$_{3,17}-20_{2,18}$, A ($E_u=$537 K), is located within the 50 au area around the protostar, and seems to have a slight extension toward the northwest. Toward the continuum peak, we also detect one emission line from CH$_2$DOH and two features of multiple CH$_3$OCHO lines. These results, in combination with previous reports, indicate that IRAS 15398$-$3359 is a source with hybrid properties showing both hot corino chemistry rich in complex organic molecules on small scales $\sim$10 au) and warm carbon-chain chemistry (WCCC) rich in carbon-chain species on large scales ($\sim$100-1000 au). A possible implication of the small emitting region is further discussed in relation to the origin of the hot corino activity.

\end{abstract}


\section{Introduction}\label{intro}

\par Ever more interstellar molecules are being found, thanks to developments in radio astronomy. More than 270 interstellar molecular species are known at present, many of which are organic molecules\footnote{https://cdms.astro.uni-koeln.de/classic/molecules}. This fact clearly reveals that a reservoir of the rich organic chemistry in the solar system had already existed in the natal interstellar clouds. Thus, exploring the chemical composition around young low-mass protostars is of particular importance to gain understanding on the molecular species inherited from interstellar clouds by planetary systems \citep[e.g.,][]{Ceccarelli et al.(2022)}.

\par In the 2000s, the chemical diversity of low-mass Class 0/I protostellar sources was revealed through sensitive millimeter-wave observations. Two distinct cases were found: the hot corino source and the warm carbon-chain chemistry (WCCC) source \citep[e.g.,][]{Cazaux et al.(2003), Sakai et al.(2008)}. In hot corino sources such as IRAS 16293$-$2422 \citep{Cazaux et al.(2003)}, NGC1333 IRAS 2A \citep{Maury et al.(2014)}, IRAS 4A \citep{Bottinelli et al.(2004)a}, and IRAS 4B \citep{Sakai et al.(2006)}, interstellar complex organic molecules (iCOMs), consisting of more than six atoms with at least one carbon atom \citep{HerbstandvanDishoeck(2009)}, are abundant around the protostar while carbon-chain molecules are generally deficient. The situation was regarded as the low-mass analog of the hot core phase seen in high-mass star forming regions. 
A hot corino is characterized by the emission of iCOMs, which are sublimated from grain mantles in a hot ($T>$100 K) region around a central source \citep[e.g.,][]{HerbstandvanDishoeck(2009),Ceccarelli et al.(2022)}.
It does not directly correspond to the disk/envelope structure, but often traces part of it \citep[e.g.,][]{Oya et al.(2016), Oya et al.(2017), Okoda et al.(2022), Maureira et al.(2022)}.
In contrast, protostellar sources rich in carbon-chain molecules (e.g., C$_4$H$_2$, \ccchh, C$_4$H, C$_5$H, CH$_3$CCH, HC$_5$N, and HC$_7$N) and deficient in iCOMs were also discovered \citep[e.g.,][]{SakaiandYamamoto(2013)}. The Class 0 protostellar source L1527 in the Taurus molecular cloud was recognized as the first WCCC source by \cite{Sakai et al.(2008)}, and \iras\ was the second discovered source \citep{Sakai et al.(2009)}. 
\par In the ALMA era, hot corinos have been found in various sources \citep[e.g.,][]{Calcutt et al.(2018), Manigand et al.(2020), Yang et al.(2020), Yang et al.(2021), van Gelder et al.(2020), Martin-Domenech et al.(2021), Nazari et al.(2021), Bouvier et al.(2022), Chahine et al.(2022), Hsu et al.(2022)}.
Hot corinos have been identified even in some sources which were hitherto classified as a WCCC source abundant in carbon-chain molecules, owing to high-resolution and high-sensitivity observations. For instance, in the low-mass protostellar sources, B335 \citep{Imai et al.(2016)} and L483 \citep{Oya et al.(2017)}, carbon-chain molecules are distributed on a 1000 au scale around the protostar, while iCOMs are apparent  on $\sim$10 au scales. Thus, WCCC and hot corino chemistry can coexist in a single source at different scales, and protostars having such a chemical property are called $'$hybrid sources$'$ \citep{Oya et al.(2017)}. Subsequently, recent observations have revealed that the low-mass protostar CB68 also has such a hybrid chemical nature \citep{Imai et al.(2022)}. 
Although the existence of intermediate sources having both the characteristics had been suggested by molecular surveys of protostellar sources \citep{Sakai et al.(2009), Lindberg et al.(2016), Higuchi et al.(2018)} with single-dish telescopes, their nature is now revealed by high-resolution observations. The origin of this chemical diversity has been discussed in terms of evolutionary and environmental effects \citep{SakaiandYamamoto(2013), Lindberg et al.(2015), Spezzano et al.(2016), Spezzano et al.(2017), Spezzano et al.(2020), Higuchi et al.(2018), Lattanzi et al.(2020), Bouvier et al.(2022)}.
Under these circumstances, it is thus important to elucidate whether a hot corino exists within prototypical WCCC sources. In this paper, we report the detection of a small hot corino in \iras, which is hitherto recognized as a WCCC source, using high-sensitivity observations by the ALMA Large program FAUST \citep[Fifty AU STudy of the chemistry in the disk/envelope system of solar-like protostars\footnote{http://faust-alma.riken.jp}:][]{Bianchi et al.(2020), Okoda et al.(2021a), Codella et al.(2021), Codella et al.(2022), Imai et al.(2022), Ohashi et al.(2022), Vastel et al.(2022)}.

\par \iras\ is a low-mass Class 0 protostellar source \citep*[\tbol$=$44 K;][]{Jorgensen et al.(2013)} located in the Lupus 1 molecular cloud \citep*[$d=$156 pc;][]{Dzib et al.(2018)}. The physical structure of this source has been studied extensively through interferometric observations, including with ALMA \citep[e.g.,][]{Jorgensen et al.(2013), Oya et al.(2014), Yildiz et al.(2015), Bjerkeli et al.(2016a), Bjerkeli et al.(2016b), Yen et al.(2017), Okoda et al.(2018), Okoda et al.(2020), Okoda et al.(2021a), Vazzano et al.(2021)}. 
\par One intriguing feature of \iras\ is that three outflows, launched toward largely different directions, are identified in spite of the apparent single protostar. In addition to the primary outflow extending from the northeast to southwest direction (P.A. 220\degr) \citep{Oya et al.(2014), Yildiz et al.(2015), Bjerkeli et al.(2016a), Yen et al.(2017), Vazzano et al.(2021)}, a relic outflow along the northwest to southeast direction (P.A. 140\degr) was discovered by \cite{Okoda et al.(2021a)}, and furthermore, an extended outflow blowing along the north to south direction (P.A. 0\degr) is visible (Sai et al. in prep.). 
The disk structure perpendicular to the direction of the primary outflow (P.A. 130\degr) was revealed with SO line emission.
The protostellar mass is estimated to be as small as 0.007$^{+0.004}_{-0.003}$\,\sm\ from the Keplerian rotation, and the disk mass between 0.006 \sm\ and 0.001 \sm\ from 1.2 mm dust continuum emission, assuming the dust temperature of 20 K and 100 K, respectively \citep{Okoda et al.(2018)}.
The envelope mass was reported to be much larger, being 0.5\,\sm\ \citep{Kristensen et al.(2012)} and 1.2\,\sm\ \citep{Jorgensen et al.(2013)}, based on the dust continuum data.
In short, \iras\ should be in the earliest stage of protostellar evolution.

\par In spite of these extensive observations, a hot corino has not been identified in this source so far, although chemical features have been discussed \citep{Jorgensen et al.(2013), Oya et al.(2014), Bjerkeli et al.(2016b), Okoda et al.(2020)}. To further explore the chemical characteristics of this source, we have conducted sensitive observations with ALMA and present the analysis in this paper. 
We present the relevant observation in Section \ref{observation}.
We show the observational results and derive the physical parameters of molecular lines in Section \ref{section3}.
We discuss the results in Section \ref{section4}, before the conclusion of this paper in Section \ref{section5}.

\section{Observations}\label{observation}

\par Single-field observations of \iras\ for two frequency setups in Band 6 were carried out between 2018 October and 2019 January as part of the ALMA Large Program FAUST. Observation parameters including calibrator sources are summarized in Table \ref{ALMA_observations}.  The field center was taken to be ($\alpha_{2000}$, $\delta_{2000}$)= (15$^{\rm h}$43$^{\rm m}$02$^{\rm s}$.242, $-$34\arcdeg 09\arcmin 06.$''$805) for both setups. In this study, we used the 12\,m array data from two array configurations (C43-5 and C43-2 for sparse and compact configurations, respectively), combining these visibility data sets in the UV plane. Molecular lines within the frequency ranges from 232 to 235 GHz (Setup 1) and from 246 to 248 GHz (Setup 2) were used in the analyses. These lines are listed in Table \ref{observed_lines}. For Setup 1 and Setup 2, the backend correlator was set to a resolution of 488 kHz (0.12\kms) with a bandwidth of 1.88 GHz and a resolution of 977 kHz (0.24 \kms) with a bandwidth of 1.88 GHz, respectively. The data were reduced in Common Astronomy Software Applications (CASA) package 5.8.1 \citep{McMullin et al.(2007)} using a modified version of the ALMA calibration pipeline and an additional in-house calibration routine to correct for the $T_{sys}$ and spectral line data normalization\footnote{See https://help.almascience.org/index.php?/Knowledgebase/Article/View/419.}. Self-calibration was carried out using line-free continuum emission for each configuration.
The approximate uncertainty for the absolute flux density scale is 10\%.
The details for self-calibration are described in \cite{Imai et al.(2022)}.
Integrated intensity map was prepared through the procedure of CLEANing the dirty images with a Briggs robustness parameter of 0.5.

\section{Results}\label{section3}
\subsection{Continuum}

\par Figure \ref{moment}(a) presents the dust continuum emission for \iras\ at 1.2\,mm, showing a centrally-peaked circular distribution. The peak position and peak intensity of the 1.2\, mm continuum are derived via a 2D Gaussian fit to the image to be ($\alpha_{2000}$, $\delta_{2000}$) = (15$^{\rm h}$43$^{\rm m}$02$^{\rm s}$.24, -34\arcdeg09\arcmin06.$''$93) and 8.69$\pm$0.32\,\mjybeam, respectively, and the synthesized beam size is 0$\farcs18\times0\farcs$16 (P.A. -84.5$\degr$). The deconvolved size of the central peak is (0\farcs13$\pm$0\farcs02$) \times ($0\farcs11$\pm$0\farcs02)  (P.A. 124\degr$\pm$65\degr) assuming a Gaussian distribution.
This size corresponds to a diameter of 20 au $\times$ 17 au. Detailed analyses of the continuum emission, including those at 1.3\,mm and 3\,mm, will be described in a separate FAUST publication.

\subsection{High Excitation Lines of \meta}\label{detection}

\par  Figures \ref{sp}(a) and (c) show the spectrum observed toward the 0.\arcsec2 ($\sim$30 au) area around the continuum peak at both frequency setups. We identify 9 \meta\ lines above the 3\,$\sigma$ noise level, where $\sigma$ means the root-mean-square (rms) noise of one channel, as listed in Table \ref{observed_lines}. The upper state energies of these lines range from 61\,K to 537\,K. The synthesized beam size for each line image is provided in Table \ref{observed_lines}, which is taken from each data cube. These are the first detections of high excitation lines of \meta\ with the upper state energies up to $\sim$500 K, toward \iras. Previously, the $K$ structure lines of the \meta\ $J=7-6$ transition were observed by \cite{Jorgensen et al.(2013)}, where the highest upper state energy is 258 K. Since their \meta\ spectrum is averaged over the 2$''$ ($\sim$300 au) area around the continuum peak, it traces a larger structure on a few 100 au scale. Rather, part of the outflow feature is likely contaminating the emission.
Note that they also reported tentative detections of CH$_3$CN and CH$_3$OCH$_3$ in the 2\arcsec\ averaged spectrum, although their origins were not discussed.
The low excitation lines of \meta, extending over a few 100 au scale or larger and tracing part of the primary and relic outflows, were also reported \citep{Okoda et al.(2020), Okoda et al.(2021a)}.

\par The spectral profiles of the three \meta\ transition lines with upper state energy higher than 400\,K (18$_{3,15}-18_{2,16}$, A, 19$_{3,16}-19_{2,17}$, A, and 20$_{3,17}-20_{2,18}$, A) observed toward the continuum peak position are depicted in Figure \ref{moment}(c).
We here employ the systemic velocity of this protostellar source of 5.5\,\kms, which is previously reported on the basis of the Keplerian rotation motion of the disk component by \citet{Okoda et al.(2018)}. These three spectra show a similar feature to each other: in particular, the blueshifted component is brighter than the redshifted component. Figure \ref{moment}(b) depicts the stacked integrated-intensity map of the blueshifted (-0.2 \kms\ to 5.0 \kms) and redshifted (5.8 \kms\ to 10.6 \kms) components for the three high-excitation lines whose upper state energies are higher than 400 K. 
The redshifted component showing only the 2.5$\sigma$  contour ($\sigma$=2.0 \mjybeam \kms) is not clear. 
The distribution of the blueshifted component is concentrated within the 50 au area around the continuum peak position, with a slight extension along the disk midplane (P.A 130\degr) toward the northwest. 
A Keplerian disk, with the radius of 40\,au in the SO emission, has been reported by \cite{Okoda et al.(2018)}. They analyzed its velocity structure and found a rotational motion blueshifted in the northwestern part as schematically shown in gray in Figure \ref{moment}(d). 
While the distribution of \meta\ is more compact than that of SO reported by \cite{Okoda et al.(2018)}, the slight extension of the blueshifted component of the \meta\ lines  toward the northwest seems consistent with the direction of the rotational motion of the disk observed in the SO line. 
Unfortunately, we do not see the redshifted component clearly in the \meta\ lines to discuss the kinematics in detail.
Nevertheless, the \meta\ lines could come from part of the disk structure.
This feature is discussed in Section \ref{vs}.

\subsection{Physical Parameters of \meta}\label{phys_meta}

\par To understand the physical conditions within the emitting region of \meta\ near the protostar, we use the observed intensities to evaluate the rotation temperature ($T_{\rm rot}=T$), the \meta\ column density ($N$), and the beam filling factor ($f$) toward the continuum peak. 
We obtain the peak intensities and the widths of the nine \meta\ lines for the blueshifted component toward the continuum peak by a Gaussian fit, as listed in Table \ref{observed_lines} (columns 8 and 10). 
Under the assumption of local thermodynamic equilibrium (LTE), the observed intensity ($T_{\rm obs}$) and the line optical depth ($\tau_{\rm line}$) are given as:
\begin{equation}\label{Tobs}
T_{\rm obs}=\frac{fc^2}{2\nu^2k_{{\rm B}}}\ \biggl[1-{\rm exp}(-\tau_{{\rm line}})\biggr]\ \biggl[B_{\nu}(T)-B_{\nu}(T_{\rm cmb})\biggr],
\end{equation}
and
\begin{equation}\label{tau}
\tau_{\rm line}=\frac{8\pi^3 S\mu^2}{3h\Delta v U(T)}\ \biggl[{\rm exp}\biggl(\frac{h\nu}{k_{\rm B}T}\biggr)-1\biggr]{\ \rm exp}\biggl(-\frac{E_u}{k_{\rm B}T}\biggr)\ N,
\end{equation}
respectively. Here, $B_{\nu}(T)$ and $B_{\nu}(T_{\rm cmb})$ are the Planck functions for the source temperature, $T$, and the cosmic microwave background temperature, $T_{\rm cmb}$ (2.7 K), respectively. Furthermore, $S$ is the line strength, $\mu$ the dipole moment responsible for the transition, $h$ the Planck constant, $\Delta v$ the velocity line width, $U(T)$ the partition function of the molecule at the source temperature $T$, $\nu$ the frequency, $E_u$ the upper-state energy, $k_{\rm B}$ the boltzmann constant, and $c$ the speed of light. The LTE condition is generally justified near the protostar ($\sim$10 au scale) where the H$_2$ density is 10$^7-$10$^8$\,cm$^{-3}$ or higher \citep{Jorgensen et al.(2013), Sakai et al.(2014a)}.  
\par
\begin{figure}[h!]
\centering
\includegraphics[scale=0.6]{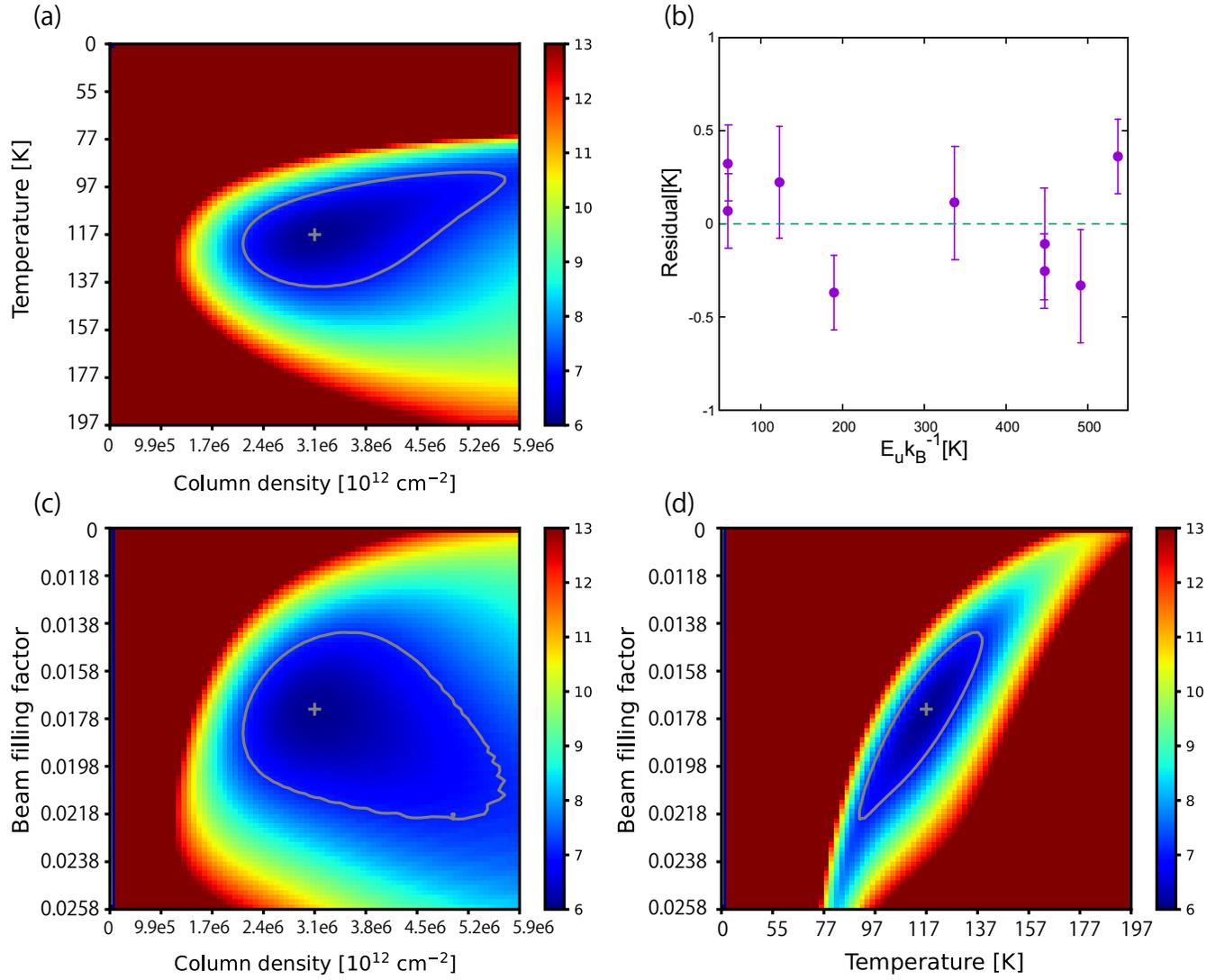}
\caption{(a, c, d)  The $\chi^2$ plots of temperature ($T$) vs. column density ($N$), beam filling factor ($f$) vs. column density ($N$), and beam filling factor ($f$) vs. temperature ($T$). Gray lines show the uncentainties of 1$\sigma$ in the $\chi^2$ analysis. (b) Residuals in the least-square fitting plotted against the upper state energies (See also Table \ref{observed_lines}). \label{fit}}
\end{figure}

 To find the best fit values of the source temperature ($T$), the column density ($N$), and the beam filling factor ($f$) using the nine \meta\ line intensities, based on the above equations (1) and (2), a non-linear least-squares fit on the observed peak intensities considering the line widths is conducted \citep{Imai et al.(2022)}, where the three parameters are optimized simultaneously.
The residuals in the fit are listed in Table \ref{observed_lines} (column 9) and shown in Figure \ref{fit}(b) in order to show the fit quality. The residuals (observed intensity minus calculated intensity for the best fit parameters) are comparable to the intensity errors, and hence, the fit is successful. 
Figures \ref{fit}(a), (c), and (d) show the $\chi^2$ plots for the $N-T$, $N-f$, and $T-f$ planes, respectively.
The minimum $\chi^2$ value for each plane corresponds to the best fit parameter values determined by the least squares fit, where the gray line represents the uncertainties (1$\sigma$).
The rotation temperature of \meta\ is thus derived to be 119$^{+20}_{-26}$\,K, where the uncertainties (1$\sigma$) are presented.
Such a high temperature is characteristic of hot corinos \citep[e.g.,][]{Ceccarelli(2004), HerbstandvanDishoeck(2009), Ceccarelli et al.(2022)}. Previously, \cite{Okoda et al.(2020)} reported the gas temperature around the protostar to be 54$\pm$2\,K, based on four \hhco\ lines observed at a resolution of 0\farcs5. Our result here shows that the innermost region is hotter. 
The column density of \meta\ and the beam filling factor are determined to be 3.2$^{+2.5}_{-1.0}\times$10$^{18}$ cm$^{-2}$ and 0.018$^{+0.005}_{-0.003}$, respectively. 
Note that the column density and the beam filling factor can be determined separately because the lines are not optically thin (Table \ref{observed_lines}).
Approximately, the relative strength of the lines provides a measure of the excitation temperature and the column density, while the actual intensities provide a measure of the beam filling factor.
The small beam filling factor suggests that the emitting region is smaller than the synthesized beam size for the \meta\ images.
If the emitting region were a round Gaussian shape on the plane of the sky, the size would be only 4 au in diameter. The size is much smaller than the disk radius of 40 au \citep{Okoda et al.(2018)}.
Considering the slightly extended distribution toward the northwester direction (Figure \ref{moment}(b)), it is likely that the \meta\ emission could arise from the thin surface layer of the disk structure.
This point is discussed further in Section \ref{vs}.
\par As the high excitation lines of $E_uk_{\rm B}^{-1}$ $>$ 400 K are included in the analysis, their excitations would be affected by infrared pumping. This effect may lead to the high rotation temperature in the analysis. To assess this possibility, we tried to conduct a similar analysis by using the 5 lines of $E_uk_{\rm B}^{-1}$ $<$ 350 K. The best fit values of the rotation temperature, the column density, and the beam filling factors obtained from the 5 lines are 107 K, 2.4$\times$10$^{18}$ cm$^{-2}$, and 0.021, respectively, which are not very different from those derived from all the lines.
However, the parameters are not well constrained due to heavy correlation among the parameters, according to the error estimation using the $\chi^2$ values.
 Hence, we assume here the LTE condition with a single rotation temperature, and report the parameters derived with all the observed \meta\ lines as the best effort. 
\par In Figures \ref{sp}(b) and (d), the spectrum of \meta\ calculated from the best-fit parameters derived in the above analysis is shown in orange. The calculated spectrum reasonably reproduces the observations within difference of up to around 3$\sigma$ noise level in the intensity. The observed spectrum are blueshifted, and hence, the systemic velocity for the calculated spectra is set to 2.9 \kms\ in Figures \ref{sp}(b) and (d).
We use the average line width of 3.4 \kms\ in the above spectrum simulation.
The clear detection of high-excitation \meta\ lines in a hot region, where the temperature is higher than 100 K within the 50 au area around the protostar, indicates the existence of a hot corino in \iras. Although this source had been regarded as a prototypical WCCC source, it also harbors a hot corino.

\subsection{\metd\ and \mf}\label{metdmf}

\par
\begin{figure}[h!]
\centering
\includegraphics[scale=0.7]{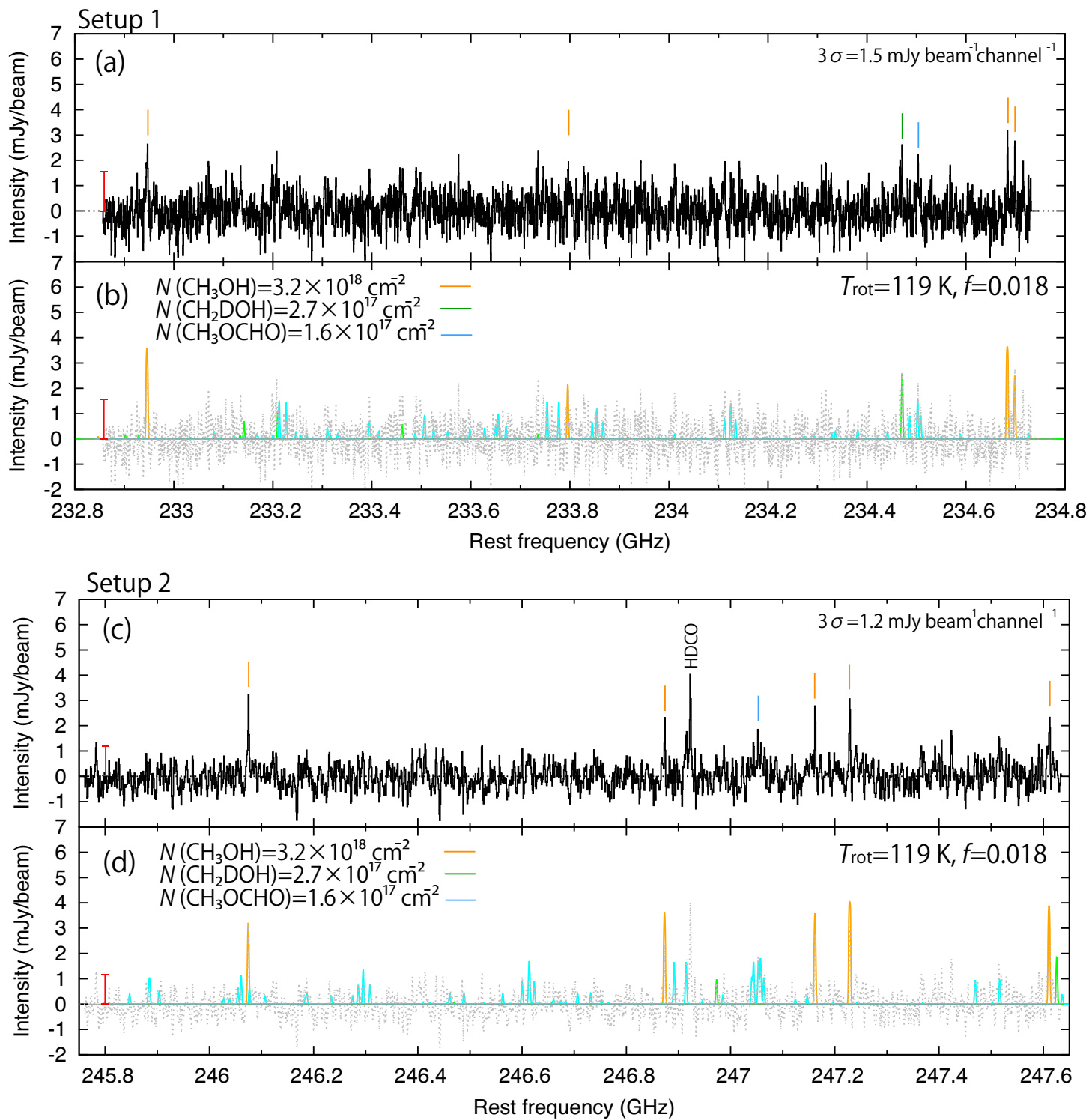}
\caption{Observed spectra of the two spectral windows extracted toward the continuum
peak in black (a,c) and gray (b,d). For this figure, the systemic velocity is set to 2.9 \kms\ for both of them to directly compare the calculated spectra to the blueshifted component of the observed spectra.
The synthesized beam size is 0$\farcs34\times0\farcs$29 for (a,b), while it is
0$\farcs20\times0\farcs$18 for (c,d). The column
density ($N$), the rotation temperature ($T_{\rm rot}$), and the beam filling factor ($f$) are derived with the LTE analysis (see text). Colors for the molecules are indicated in the upper left corners of (b) and (d). The line width is assumed to be 3.4 \kms\ for all lines. The red vertical bars on the left of each panel show the 3$\sigma$ (three times the rms noise per one channel) level. \label{sp}}
\end{figure}

 We here discuss detections of the \metd\ and \mf\ lines in the observed spectrum (Figure \ref{sp}) based on the rms noise of the integrated intensity. 
In Figure \ref{sp}(a), we find a faint line of \metd\ (8$_{2,6}-8_{1,7}$, e$_0$: $E_uk_{\rm B}^{-1}=$94 K) at 234.471\,GHz, whose integrated intensity is 13 \mjybeam\kms.
Since the rms noise level of the integrated intensity is 2.7 \mjybeam\kms, the line is detected with the confidence level of 4.8$\sigma$. This line has the upper state energy of 94 K, and has been detected in hot corinos around other protostellar sources by the FAUST program: e.g., L1551 \citep{Bianchi et al.(2020)} and CB68 \citep{Imai et al.(2022)}. 
In CB68, another \metd\ line at 247.6257463\,GHz, (3$_{2,1}-3_{1,2}$, e$_0$: $E_uk_{\rm B}^{-1}=$29 K) was detected as well.
However, it is not seen with the 3$\sigma$ confidence level in \iras, although it is expected to be deteted in the calculation (Figure \ref{sp} (d)).
 We do not fully account for this non-detection, but here note two possibilities.
 Since this line is quite low-excitation, it would be seriously affected by the absorption due to the extended foreground gas.
 Alternatively, the line is very close to the band edge of the correlator, which would provide a less accurate calibration.
\par The column density of \metd\ is evaluated by using equations (1) and (2). Since only one line is available for \metd, the rotation temperature and the beam filling factor are assumed to be the same as those determined for \meta, and the line width is employed for the average value of the \meta\ lines (3.4 \kms). 
Then, the column density of \metd\ is derived to be (1.8-4.3)$\times$10$^{17}$ cm$^{-2}$, where the errors of the rotation temperature and beam filling factor derived from the \meta\ lines are considered in the error estimation.
The \metd/\meta\ ratio is evaluated to be 0.03-0.20.
\par The presence of \mf\ is expected, based on the physical parameters derived from the \meta\ lines. Although most of the possible features are weak in the observed spectrum of Figure \ref{sp}, the lines at 234.5\,GHz and a cluster of lines at 247.1\,GHz seem evident.
For the 234.5\,GHz features, two lines of 19$_{9,11}-18_{9,10}$, A, and 19$_{9,10}-18_{9,9}$, A are blended, and the peak intensity integrated for these lines is 10 \mjybeam\kms.
Since the rms noise for the same velocity range is 3.0 \mjybeam\kms, the feature is detected with the 3.3$\sigma$ confidence level.
For the 247.1\,GHz feature, the six lines of 10$_{5,6}-9_{4,6}$, E, 21$_{3,19}-20_{3,18}$, A and E, 20$_{9,12}-19_{9,11}$, A and E, and 20$_{9,11}-19_{9,10}$, A, are blended, where the intensity over these lines is 40 \mjybeam\kms.
Since the rms noise is 8.0 \mjybeam\kms, the feature is detected with the 5$\sigma$ confidence level.
Hence, we report the possible detection of \mf\ in this source. Since these are high excitation lines with $E_uk_{\rm B}^{-1}$ $>$ 100 K, they would most likely come from the hot corino. We estimate the column density of \mf\ by using equations (1) and (2), to be (1.1-3.2)$\times$10$^{17}$cm$^{-2}$ and the \mf/\meta\ ratio  to be 0.02-0.15. Here, the rotation temperature and the beam filling factor are assumed to be the same as those derived for \meta.
We take into account the calculated errors of the rotation temperature and the beam filling factor derived from the \meta\ lines for the error estimates. 

\section{Discussion}\label{section4}
\subsection{Hybrid Chemical Nature of \iras}

\par As mentioned in Section \ref{intro}, \iras\ was recognized as the second WCCC source by \cite{Sakai et al.(2009)}. 
The authors detected various carbon-chain molecules toward the protostar in the 3\,mm band using the Nobeyama 45\,m telescope and the Mopra 22\,m telescope and revealed the CCH ($N=4-$3) emission concentrated around the protostar with single-dish observations from the ASTE 10\,m telescope.  Later, the rotating envelope structure just outside the Keplerian disk and the outflow along the northwest to southeast axis were reported for the CCH ($N=4-$3 and $N=3-$2) lines \citep{Oya et al.(2014), Okoda et al.(2018)}. These features are very similar to those found in the first WCCC source L1527 \citep{Sakai et al.(2008), Sakai et al.(2014a), Sakai et al.(2014b)}. 
 In the representative WCCC sources, the present detection of hot corino activity gives an important clue to our understanding of the chemical structures and the chemical diversity of low-mass protostellar sources.   
\par The existence of a hot corino clearly indicates that \iras\ has a hybrid chemical nature, as found in the other low-mass protostellar sources, L483, B335, and CB68 \citep{Oya et al.(2017), Imai et al.(2016), Imai et al.(2022)}. Among these sources with hybrid properties, CB68, which is located in Ophiuchus molecular cloud complex \citep[$d=$144$\pm$7 pc:][]{Zucker et al.(2019)}, was also observed by the FAUST program with the same spectral setting, and hence, we can directly compare the spectral features between CB68 and \iras. Although the bolometric luminosity of \iras\ is 1.8 \sL\ \citep{Jorgensen et al.(2013)}, which is even higher than that of CB68 \citep[\rm 0.86 \sL:][]{Launhardt et al.(2013)}, the intensity of the highest excitation line of \meta\ (20$_{3,17}-20_{2,18}$, A) observed in \iras\ is $\sim$3\,mJy\,beam$^{-1}$ and weaker than those in CB68 typically by a factor of $\sim$5. In both sources, the \meta\ lines are optically thick and not well-resolved spatially. Therefore, this difference means a smaller size and/or lower temperature of the hot corino of \iras\ in comparison with that of CB68. \iras\ is in an early stage of protostellar evolution \citep{Okoda et al.(2018)} with a lower protostellar mass, and it seems likely that the hot corino activity has not yet developed well. 
\par According to the models by \cite{van Gelder et al.(2022)} and \cite{Nazari et al.(2022)}, the appearance of \meta\ emission can be affected by small-scale structures within protostellar sources, in particular the presence of a disk. A similar suggestion is also given by the chemical network calculation \citep{Aikawa et al.(2020)}.  In the presence of a disk, \meta\ is depleted onto dust grains and the emission from the remaining \meta\ emission in the gas phase can be shielded by the high dust opacity \citep[e.g.,][]{Sahu et al.(2019), De Simone et al.(2020)}. \iras\ has a small disk structure of $\sim$40 au in radius \citep{Okoda et al.(2018)}, while the disk size of CB68 is smaller than 30 au in radius \citep{Imai et al.(2022)}.
Thus, neither \iras\ nor CB68 have a disk structure larger than 50 au, which \cite{van Gelder et al.(2022)} employed to be the threshold size for defining small disks.  
The peak intensities of the dust emission are similar to each other: 5.73 K (CB68) \citep{Imai et al.(2022)} and 4.39 K (\iras) at a resolution of $\sim$0.\arcsec2.
Nevertheless, the intensity of the \meta\ emission is different by a factor of 5 between them, as mentioned above. 
Furthermore, the luminosity of \iras\ is twice that of CB68 (Table \ref{abundance}).
These features would suggest that the appearance of \meta\ emission is not solely dependent on the current physical disk structure nor the source luminosity. 
\par The observed hybrid chemical nature is predicted by chemical network calculations.
According to \cite{Aikawa et al.(2008), Aikawa et al.(2020)}, WCCC occurs at the 1000 au scale ($\sim$25\,K) due to sublimation of CH$_4$. On the other hand,  hot corino chemistry occurs within the 100 au scale ($\gtrsim$ 100\,K) due to sublimation of H$_2$O and iCOMs. Our result supports the notion indicated by \cite{Oya(2020)} that a hybrid chemical nature is the common occurrence in the chemical structure of protostellar cores. The relative appearance of WCCC and hot corino activity would thus depend on both the grain mantle composition as well as the physical conditions around the protostar \citep[e.g.,][]{SakaiandYamamoto(2013), Lindberg et al.(2015), Spezzano et al.(2016), Higuchi et al.(2018)}. In this relation, different chemical compositions at the prestellar phase would impose different chemical features at the protostellar phase. 
In some prestellar cores including L1544, the chemical differentiation in a core was reported \citep{Spezzano et al.(2016),Spezzano et al.(2017),Spezzano et al.(2020), Soma et al.(2015), Soma et al.(2018)}, showing that there are regions rich in carbon-chain molecules on one side and organic molecules like \meta\ are distributed around a different part of the core.
Such a feature is proposed to be caused for instance by the differential illumination from the interstellar radiation field.
These observational results give an expectation for the presence of WCCC and hot corino chemistry as a mixing of ice composition during its evolution and in the future protostellar envelope and disk.
Moreover, recent studies indicate that the presence of streamers could modify the local abundance of carbon bearing molecules \citep{Pineda et al.(2020), Pineda et al.(2022), Valdivia-Mena et al.(2022)}.
\par Apparently, high resolution and high sensitivity observations of more protostellar sources are necessary \citep[e.g.,][]{Yang et al.(2021)}.
In particullar, it will be interesting to examine whether any hot corino activity can be seen around the first discovered WCCC source, L1527. 
It should also be noted that some sources show only a hot corino chemistry, such as NGC 1333 IRAS2A and IRAS 16293-2422 \citep[e.g.,][]{Calcutt et al.(2018), Taquet et al.(2015), Taquet et al.(2019), Jorgensen et al.(2018), Manigand et al.(2020)}.
It will be important to study a WCCC feature for these sources as well.

\subsection{Comparison with Other Hot Corinos}

\par We summarize and compare the abundances of \metd\ and \mf\ relative to \meta\ among low-mass protostars having a hot corino  in Table \ref{abundance}.
The ratios in \iras\ are based on the assumption that the emitting regions of \metd\ and \mf\ are the same as that of \meta, as mentioned in Section \ref{metdmf}. The three hybrid sources (B335, L483, and CB68) and three hot corino sources (L1551 IRS5, NGC 1333 IRAS2A, and IRAS 16293-2422B) are selected for this purpose.
Although many hot corinos are observed, these sources are appropriate samples for comparison, because they are observed at a high resolution.
Comparison among larger samples was done by \cite{van Gelder et al.(2020)}, \cite{Hsu et al.(2022)}, and \cite{Imai et al.(2022)}.
While the \metd/\meta\ ratio in \iras\ (0.03-0.20) has a large uncertainty, it indicates moderate deuterium fractionation comparable to those for the other protostellar sources (a few \%) except for B335.
The high ratio for B335 seems to originate from a very small beam size in the observation, as pointed out by \cite{Okoda et al.(2022)}.

\par The \mf/\meta\ ratio in \iras\ is derived to be 0.02-0.15. This is higher than those reported for L483 (0.0076) \citep{Jacobsen et al.(2019)}, NGC 1333 IRAS2A (0.016$^{+0.012}_{-0.007}$) \citep{Taquet et al.(2015)}. The ratios for the other sources, CB68 (0.09$^{+0.07}_{-0.03}$) \citep{Imai et al.(2022)} and L1551 IRS5 (0.033$\pm$0.002) \citep{Bianchi et al.(2020)}, and IRAS 16293-2422B (0.026$^{+0.008}_{-0.007}$) \citep{Jorgensen et al.(2018)}, are within the range of the ratios in \iras. The observed \mf/\meta\ ratio implies that a substantial amount of \mf\ is present in \iras.

\par The bolometric luminosity of \iras\ is 1.8 \sL \citep{Jorgensen et al.(2013)}, which is similar to that of B335 and twice that of CB68. Even higher bolometric luminosities are reported for L483 \citep[13 \sL:][]{Shirley et al.(2000)} and L1551 IRS5 \citep[30-40 \sL:][]{Liseau et al.(2005)}. Nevertheless, the \metd/\meta\ and \mf/\meta\ ratios in L483 and L1551 IRS5 are similar to the other sources.
This has also been revealed by \cite{van Gelder et al.(2020)} and \cite{Imai et al.(2022)}.
Now, we find that \iras\ is not an exception in these ratios.

\subsection{Small Beam Filling Factor vs. Slight Extension toward the Northwest}\label{vs}

\par
\begin{figure}[h!]
\centering
\includegraphics[scale=0.42]{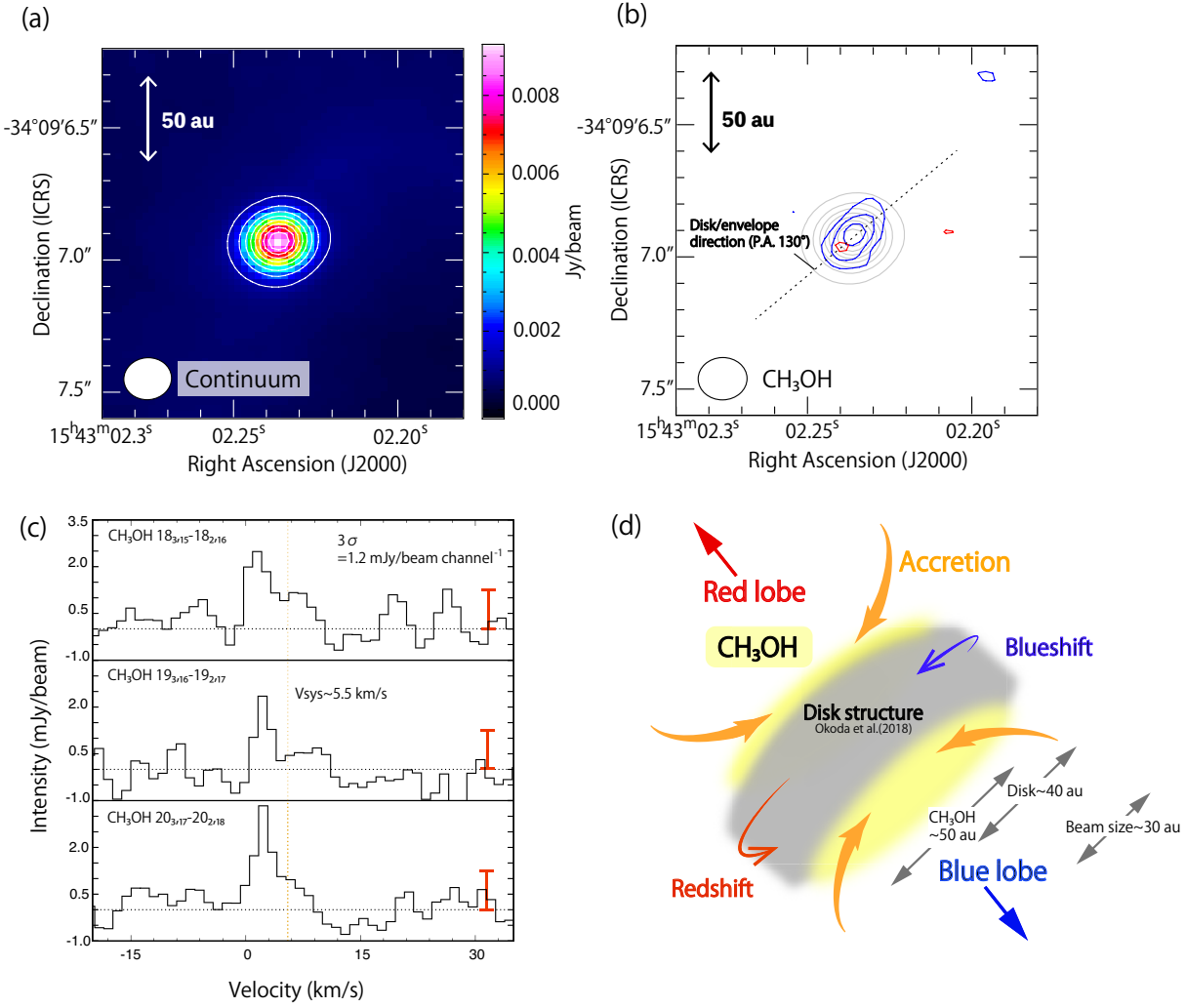}
\caption{(a) 1.2 mm continuum emission in the Setup 2 observation. Contour levels are every 10$\sigma$ from 3$\sigma$, where $\sigma$ is 0.1\mjybeam.
The white ellipse at the bottom-left corner shows the synthesized beam size. (b) Stacked moment 0 maps of the blueshifted and redshifted components for the \meta\ (18$_{3,15}-18_{2,16}$, A, 19$_{3,16}-19_{2,17}$, A, and 20$_{3,17}-20_{2,18}$, A) lines ($E_u>$400 K). The blueshifted and redshifted components are shown in blue and red contours, respectively. The contour levels for the blueshifted component are every $\sigma$ (the rms noise of the intensity map) from 3$\sigma$ and that for the redshifted one is 2.5$\sigma$, where $\sigma$ is 2.0 \mjybeam\kms. The integrated velocity range of the blueshifted component is from -0.2 \kms\ to 5.0 \kms\ ($V_{sys}$$\sim$5.5 \kms), and that of the unclear redshifted component from 5.8 \kms\ to 10.6 \kms. The open ellipse at the bottom-left corner shows the synthesized beam size of the \meta\ observations. (c) Spectra of the three \meta\ lines toward the continuum peak. The red vertical bars show the 3$\sigma$ (three times the rms noise per one channel) level. The horizontal and vertical dotted lines show the zero-level intensity and $V_{\rm sys}$, respectively. (d) Schematic illustration of the possible \meta\ distribution in the disk structure. Yellow shows the thin surface layers of the disk structure where the \meta\ emission arises (See Section \ref{vs}). \label{moment}}
\end{figure}

 In Section \ref{phys_meta}, we find a small beam-filling factor (0.018$\pm$0.001), based on an analysis of the \meta\ line intensities, while the emission is slightly extended beyond the synthesized beam size toward the northwest (Figure \ref{moment}(b)).
The beam filling factor approximately represents the ratio of the emitting area on the plane of the sky to the beam size.
As noted in Section \ref{phys_meta}, if the emitting area were a round Gaussian shape, its diameter would be $\sim$4\,au based on the beam filling factor. 
Hence, these two results look contradictory at first glance, and yet can be reconciled.
\par Even for the small beam filling factor, an extend feature is possible. 
The emission of the high-excitation lines of \meta\ does not necessarily come from a single round-shaped structure. In principle, it might come from very small-scale features distributed over the area slightly larger than the synthesized beam size. 
One possibility which we speculate is that the \meta\ emission could arise from the thin surface layer of the disk structure, as indicated by the yellow part in the schematic illustration of Figure \ref{moment}(d). 
Such a distribution of \meta\ is indeed reported for the low-mass Class 0 protostellar source HH212 by \cite{Lee et al.(2017), Lee et al.(2018)}. These authors also found a similar distribution for the \metd, CH$_3$SH, and \nhhcho\ emission in HH212. The HH212 emission is seen from both surfaces of the disk, sandwiching the continuum distribution, presenting a `hamburger shape' structure. 
Since rotational motion is evident, \cite{Lee et al.(2017)} conclude that the HH212 emission traces the disk structure with part of it arising from the disk atmosphere produced by the gas accreting onto the disk. 
This picture could also be the case for \iras, which has a disk/envelope system at an almost edge-on configuration \citep[$i\sim$20\degr:][]{Oya et al.(2014)}, as HH212. If only the thin surface layer harboring iCOMs is visible, the beam filling factor will be small. In this case, the slight extension to the northwest (by $\sim$20 au) seems possible, as the disk radius is reported to be 40\,au by \cite{Okoda et al.(2018)}. As shown in Figure \ref{moment}(b), the blueshifted velocity and the slight extension toward the northwestern side in the \meta\ lines is consistent with the direction of the rotational motion of the disk reported previously \citep{Okoda et al.(2018)}.
\par On the other hand, weaker redshifted emission (Figure \ref{moment}(b)) is puzzling. Indeed, the redshifted components are hardly seen in the spectral profiles of Figure \ref{moment}(c).
The absorption at the systemic velocity (5.5 \kms) might be caused by the absorption due to the foreground gas, but an inverse P-cygni profile could more reasonably explain the asymmetric spectral profiles of \meta\ lines (Figure \ref{moment}(c)).
If this were the case, the \meta\ emitting gas would be infalling on the disk structure, as shown in Figure \ref{moment}(d).
Needless to say, this is not an only possibility.
If the \meta\ emission comes from the launching point of the primary outflow, its redshifted part could be shielded by dust of the disk midplane.
It should also be noted that the weak redshifted emission could also originate from asymmetric distribution of the \meta\ emitting gas.
In any case, the present observation cannot allow us to discuss further detailed kinematics of the \meta\ emitting region.
\par COMs can possibly be liberated from the dust grains by the accretion shock heating and/or the protostellar heating on a few au scale.
In the former case, we expect that gas obliquely accreting with respect to the disk midplane would in principle cause an accretion shock heating the surface \citep{Miura et al.(2017)}. 
Taking into account for the small protostellar mass \citep[0.007$^{+0.004}_{-0.003}$ \sm:][]{Okoda et al.(2018)} and the small size of the hot corino, the accretion velocity at $r=$ 2 au can be 2.4 \kms\ in the free fall case.
This velocity is enough for sublimation of ice mantles in the case of the high H$_2$ density ($n>$10$^9$ cm$^{-3}$) \citep[]{Miura et al.(2017)}.
In fact, the blueshifted component has the velocity of $\sim$3 \kms, as shown in Figure \ref{moment}(c).
The gas on the disk surface up to 10 au in radius can also efficiently be heated by illumination from the protostar, according to the model by \cite{Walsh et al.(2010)}. 
Note that photodesorption of \meta\ is inefficient, as shown by the laboratory experiments \citep{Bertin et al.(2016), Cruz-Diaz et al.(2016)}.
Thus, thermal desorption in the surface layer should be considered here.
A combination of the accretion and irradiation effects likely raises the temperature high enough to sublimate ice mantles ($>$ 100 K) and/or sputter surface molecules from dust grains, resulting in the observed hot corino activity. 
Figure \ref{moment}(d) illustrates a schematic sketch of the \meta\ emitting region on the disk thin surface caused by the accretion shock with the yellow part and the orange arrows.
To examine this hypothesis, we need higher angular resolution observations to resolve the vertical structure of the disk.

\section{Summary}\label{section5}

To understand the chemical structure on a 50 au scale around the low-mass protostar  \iras, we have conducted sensitive observations with the synthesized beam size of $\sim$30 au as part of the ALMA Large Program FAUST. The main results are summarized below.

\par 1. We detect 9 high excitation lines of \meta\ toward the continuum peak. We derive a rotation temperature of 119$^{+20}_{-26}$\,K, a column density of 3.2$^{+2.5}_{-1.0}\times$10$^{18}$cm$^{-2}$, and a beam filling factor of 0.018$^{+0.005}_{-0.003}$ using a least-squares fit on the observed intensities under assumption of the LTE conditions. Such a high temperature is a characteristic feature of hot corinos, and therefore, this study reveals the existence of a small hot corino in \iras. This source is thus found to be a source with hybrid properties like B335, L483, and CB68, with a WCCC nature on large 100-1000 au scales and a hot corino on much smaller scales. The present detection of a hot corino in this prototypical WCCC source has important implications for the chemical diversity of protostellar sources.

\par 2. We detect faint \metd\ emission as well as possible features from \mf\ in the ALMA spectra toward the continuum peak. We roughly estimate the column densities of \metd\ and \mf\ to be (1.8-4.3)$\times$10$^{17}$cm$^{-2}$ and (1.1-3.2)$\times$10$^{17}$cm$^{-2}$, respectively, assuming the rotation temperature and the beam filling factor determined for \meta.

\par 3. The \metd/\meta\ and \mf/\meta\ ratios in \iras\ are derived to be 0.03-0.20 and 0.02-0.15, respectively. 
The \metd/\meta\ ratio in \iras\ indicates that the deuterium fractionation level is comparable to those in the other low-mass protostars harboring a hot corino except for B335. As well, the \mf/\meta\ ratio in \iras\ is found to be similar to those in the sources CB68 \citep{Imai et al.(2022)}, L1551 IRS5 \citep{Bianchi et al.(2020)}, and IRAS 16293-2422B \citep{Jorgensen et al.(2018)}.

\par 4. The small derived beam filling factor, 0.018$^{+0.005}_{-0.003}$ derived from the analysis of the \meta\ line intensities indicates a hot corino within 4 au in diameter around the protostar, assuming a round shape, while the blueshifted distribution of the \meta\ lines with upper state energy higher than 400\,K slightly extends toward the northwest ($\sim$30 au). We speculate that the \meta\ emission comes from the thin surface of the disk structure, as reported for the low-mass protostellar source HH212 by \cite{Lee et al.(2017),Lee et al.(2018)}. Understanding the structure more clearly will require future higher-angular resolution observations.

\acknowledgments

\par The authors thank the anonymous reviewer for valuable comments.
The authors also thank the FAUST team members for fruitful discussions.
This paper makes use of the following ALMA data set: ADS/JAO.ALMA\# 2018.1.01205.L (PI: Satoshi Yamamoto). ALMA is a partnership of the ESO (representing its member states), the NSF (USA) and NINS (Japan), together with the NRC (Canada) and the NSC and ASIAA (Taiwan), in cooperation with the Republic of Chile. The Joint ALMA Observatory is operated by the ESO, the AUI/NRAO, and the NAOJ. The authors thank the ALMA staff for their excellent support. The National Radio Astronomy Observatory is a facility of the National Science Foundation operated under cooperative agreement by Associated Universities, Inc.
 This project is also supported by a Grant-in-Aid from Japan Society for the Promotion of Science (KAKENHI: Nos. 18H05222, 19H05069, 19K14753, 22K20390).
Y. Okoda thanks Japan Society for the Promotion of Science (JSPS) and RIKEN Special Postdoctoral Researcher Program (Fellowships) for financial support. D. Johnstone is supported by NRC Canada and by an NSERC Discovery Grant.

{}

\begin{table}[ht]
\caption{\centering{Observation Parameters$^a$}\label{ALMA_observations}}
\scalebox{0.7}{
\begin{tabular}{cccccc}
\hline
& \multicolumn{2}{c}{Setup 1} &   \multicolumn{2}{c}{Setup 2} &  \\
\hline
Parameter & (C43-5) & (C43-2) & (C43-4) & (C43-1) &  \\
Observation date & 2018 Nov. 23 & 2019 Jan. 6 & 2018 Nov. 25 &  2019 Jan. 14\\
Time on source (min) & 47.1 & 12.63 & 33.92 & 11.63 \\
Number antennas & 45 & 47 & 43 & 48 \\
Primary beamwidth (arcsec) & 26.7 & 26.7 & 23.7 & 23.7 \\
Frequency range (GHz)&\multicolumn{2}{c}{232-235}&\multicolumn{2}{c}{246-248}\\
Continuum bandwidth (GHz) & 1.875 & 1.875 & 1.875 & 1.875 \\
Baseline range (m) & \multicolumn{2}{c}{15.1-1397.8} &  \multicolumn{2}{c}{15.0-1397.8}  \\
Bandpass calibrator & J1427-4206& J1427-4206 & J1427-4206 & J1517-2422 \\
Phase calibrator & J1626-2951& J1517-2422& J1626-2951& J1626-2951\\
Flux calibrator & J1427-4206& J1427-4206& J1427-4206 & J1517-2422 \\
Pointing calibrator & J1650-2943 & J1427-4206 & J1650-2943, J1427-4206& J1517-2422 \\
rms (mJy beam$^{-1}$ channel$^{-1}$) & 1.8 & 4.4 & 3.3 & 5.5 \\

\hline
\end{tabular}}
\begin{flushleft}
\tablecomments{
$^a$ These observations are conducted with the Band 6 receiver of ALMA. 
}
\end{flushleft}
\end{table}

\begin{longrotatetable}
\begin{table}[ht]
\centering
\caption{\centering{Observed Molecular Lines$^a$} \label{observed_lines}}
\scalebox{0.7}{
\begin{tabular}{cccccccccccc}
\hline \hline
Setup$^b$ & Molecule&Transition & Frequency & $S \mu^2$ & $E_{\rm u}$$k_{\rm B}^{-1}$ & Synthesized beam size&Intensity$^c$&Residual$^d$&$\Delta v$$^c$&$v_{\rm center}$$^c$&$\tau$\\
&&&(GHz)&($D^2$)&(\rm K)&&(K) &(K)&(\kms) &(\kms)&\\
\hline
1 & CH$_3$OH & 10$_{3,7}-11_{2,9}$, E & 232.9457970 & 12.142 & 190 & 0$''$339$\times$0$''$289 (P.A. 74.3$^\circ$) & 1.9 (0.2) &-0.37&4.9 (0.7)&2.8 (0.3)&3.0\\
1 & CH$_3$OH & 18$_{3,15}-17_{4,14}$, A & 233.7956660 & 21.868 & 447  & 0$''$339$\times$0$''$289 (P.A. 74.3$^\circ$) & 1.2 (0.3)&-0.11&2.8 (0.7)&1.7 (0.3)&1.1\\
1 & CH$_3$OH & 4$_{2,3}-5_{1,4}$, A & 234.6833700& 4.4844 & 61  & 0$''$342$\times$0$''$291 (P.A. 74.1$^\circ$) & 2.0 (0.2)&0.07&2.4 (0.2)&2.3 (0.1)&6.8\\
1 & CH$_3$OH & 5$_{4,2}-6_{3,3}$, E & 234.6985190 & 1.8536 & 123  & 0$''$341$\times$0$''$290 (P.A. 74.3$^\circ$) & 2.0 (0.3)&0.22&1.9 (0.2)&2.7 (0.1)&2.1\\
2 & CH$_3$OH & 20$_{3,17}-20_{2,18}$, A & 246.0746050 & 78.029 & 537  & 0$''$202$\times$0$''$182 (P.A. 85.1$^\circ$) & 1.8 (0.2) &0.40&4.2 (0.5)&2.5 (0.2)&1.3\\
2 & CH$_3$OH & 19$_{3,16}-19_{2,17}$, A & 246.8733010 & 73.682 & 491  & 0$''$201$\times$0$''$181 (P.A. -88.0$^\circ$) & 1.6 (0.3)&-0.30&2.1 (0.5)&2.4 (0.2)&3.6\\
2 & CH$_3$OH & 16$_{2,15}-15_{3,13}$, E & 247.1619500 & 19.318 & 338  & 0$''$200$\times$0$''$180 (P.A. -88.4$^\circ$) & 2.0 (0.3)&0.10 &2.4 (0.5)&2.6 (0.2)&3.0\\
2 & CH$_3$OH & 4$_{2,2}-5_{1,5}$, A & 247.2285870 & 4.344 & 61  & 0$''$202$\times$0$''$182 (P.A. -85.0$^\circ$) & 2.3 (0.2)&0.30&3.3 (0.5)&2.9 (0.2)&5.1\\
2 & CH$_3$OH & 18$_{3,15}-18_{2,16}$, A & 247.6109180 & 69.431 & 447  & 0$''$200$\times$0$''$181 (P.A. -88.3$^\circ$) & 1.3 (0.2)&-0.30&6.8 (1.4)&2.9 (0.6)&1.5\\
\hline
1 & CH$_2$DOH & 8$_{2,6}-8_{1,7}$, e$_o$ & 234.4710333 & 9.550 & 94 & 0$''$341$\times$0$''$290 (P.A. 74.3$^\circ$) &0.65$^d$&-& -& -\\
\hline
1 & CH$_3$OCHO & 19$_{9,11}-18_{9,10}$, A and 19$_{9,10}-18_{9,9}$, A & 234.5022407&157.106& 166 & 0$''$341$\times$0$''$290 (P.A. 74.3$^\circ$)  & 0.52$^d$& - & -& -\\
%
2 & CH$_3$OCHO & 21$_{3,19}-20_{3,18}$, E &247.0441460&108.057&140& 0$''$200$\times$0$''$181 (P.A. -88.4$^\circ$) &1.3$^e$&-& -& -\\
2 & CH$_3$OCHO & 10$_{5,6}-9_{4,6}$, E &247.0526577 &4.433 & 236 & 0$''$200$\times$0$''$181 (P.A. -88.4$^\circ$) &$1.3^e$&-& -& -\\
2 & CH$_3$OCHO & 21$_{3,19}-20_{3,18}$, A &247.0534530 &108.070  & 140  & 0$''$200$\times$0$''$181 (P.A. -88.4$^\circ$) &1.3$^e$&-& -& -\\
2 & CH$_3$OCHO & 20$_{9,12}-19_{9,11}$, A &247.0572591&85.039&178& 0$''$200$\times$0$''$181 (P.A. -88.4$^\circ$) &1.3$^e$&-& -& -\\
2 & CH$_3$OCHO & 20$_{9,11}-19_{9,10}$, A &247.0577373&85.039&178& 0$''$200$\times$0$''$181 (P.A. -88.4$^\circ$) &1.3$^e$&-& -& -\\
2 & CH$_3$OCHO & 20$_{9,12}-19_{9,11}$, E &247.0636620&85.033&178& 0$''$200$\times$0$''$181 (P.A. -88.4$^\circ$) &1.3$^e$&-& -& -\\
\hline
\end{tabular}
}
\begin{flushleft}
\tablecomments{
$^a$ Spectroscopic line parameters are taken from CDMS \citep{Endres et al.(2016)} and JPL \citep{Pickett et al.(1998)}.
$^b$ Observation Setups (Section \ref{observation}).
$^c$ Observed intensities, line widths, and center velocities for the blueshifted component except for \metd\ and \mf.
The intensities of \meta\ in Setup 1 are normalized for the synthesized beam size of 0.\arcsec20$\times$0.\arcsec18.
The errors are in parentheses.
$^d$ The residuals in the least-squares analysis for the \meta\ lines (Section \ref{phys_meta}).
$^e$ Gaussian fit is not conducted. The peak intensity is shown here.}
\end{flushleft}
\end{table}
\end{longrotatetable}

\begin{longrotatetable}
\begin{table}[ht]
\caption{\centering{Column Densities and Relative Abundances to \meta} \label{abundance}}
\scalebox{0.8}{
\begin{tabular}{ccccccccccc}
\hline \hline
& This work & \multicolumn{3}{c}{Hybrid source (WCCC+Hot corino)}  &  \multicolumn{3}{c}{Hot corino} \\
Source & \iras & B335$^a$ & L483$^b$ & CB68$^c$& L1551 IRS5$^d$&NGC 1333 IRAS2A & IRAS 16293-2422B$^e$ \\
\hline
$L_{\rm bol}$ (\sL) & 1.8 & 1.6& 13 & 0.86 &30-40&-&-\\
Resolution$^f$  (au)& 30-50 &5& 28 & 80 &50& 300-600&60 \\
$N$(\meta) (10$^{18}$ cm$^{-2}$)&3.2$^{+2.5}_{-1.0}$&8.5$\pm$2.3&17&2.7$\pm$1.0&$\geq$10&5.0$^{+2.9}_{-1.8}$&10$\pm$2\\
$N$(\metd) (10$^{18}$ cm$^{-2}$)&0.18-0.43&2.5$\pm$1.4&0.4&0.15&$\geq$0.5&0.29$^{+0.12}_{-0.06}$&0.71$\pm$0.14\\
$N$(\mf) (10$^{18}$ cm$^{-2}$)&0.11-0.32&-&0.13&0.23$\pm$0.03&0.33$\pm$0.02&0.079$^{+0.046}_{-0.016}$&0.26$\pm$0.05\\
$\lbrack$\metd$\rbrack$/$\lbrack$\meta$\rbrack$ & 0.03-0.20 & 0.29$\pm0.2$ & 0.024 & 0.06$^{+0.04}_{-0.02}$   &$\geq$0.05$^g$&0.058$^{+0.084}_{-0.032}$$^h$& 0.07$\pm0.02$\\
$\lbrack$\mf$\rbrack$/$\lbrack$\meta$\rbrack$ & 0.02-0.15 & -& 0.0076 & 0.09$^{+0.07}_{-0.03}$  &0.033$\pm$0.002&0.016$^{+0.012}_{-0.007}$$^i$& 0.026$^{+0.008}_{-0.007}$\\
\hline
\end{tabular}
}
\begin{flushleft}
\tablecomments{
$^a$ \cite{Okoda et al.(2022)}.
$^b$ \cite{Jacobsen et al.(2019)}.
$^c$ \cite{Imai et al.(2022)}. 
$^d$\cite{Bianchi et al.(2020)}.
$^e$\cite{Jorgensen et al.(2018)}.\\
$^f$ Angular resolution taken from each observation data.
$^g$ The lower limits to the column densities of \metd\ and \meta\ are estimated.
$^h$\cite{Taquet et al.(2019)}.
$^i$\cite{Taquet et al.(2015)}.
}
\end{flushleft}
\end{table}
\end{longrotatetable}

\clearpage

\clearpage

\end{document}